\documentclass{elsart1p}

\usepackage{graphics}
\usepackage{graphicx}
\usepackage{epsfig}

\usepackage{amssymb}

\begin{document}

\begin{frontmatter}


\title{Overview of nucleon structure studies \thanksref{label1}}
\thanks[label1]{This work is supported in part by DOE grant
DE-FG02-04ER41302 and contract DE-AC05-06OR23177 under
which Jefferson Science Associates operates the Jefferson Lab. }
\author{Marc Vanderhaeghen}
\ead{marcvdh@jlab.org}

\address{Department of Physics, 
College of William and Mary, Williamsburg, Virginia 23187, USA}
\address{Thomas Jefferson National Accelerator Facility, Newport News, 
VA 23606, USA}

\begin{abstract}
A brief overview of the recent activity in the measurement of the 
elastic electromagnetic proton and neutron form factors is presented. 
It is discussed how the quality of the data has been 
greatly improved by performing double polarization experiments, 
and the role of two-photon exchange processes will be highlighted.  
The spatial information on the quark charge distribibutions 
in the nucleon resulting
from the form factors measurements will be discussed, 
as well as the steady rate of improvements made in the lattice QCD 
calculations. It is discussed how generalized parton distributions 
have emerged as a unifying theme in hadron physics 
linking the spatial densities extracted from form factors 
with the quark momentum distribution information residing 
in quark structure functions. The recent progress in the electromagnetic 
excitation of the $\Delta(1232)$ resonance will also briefly be discussed. 
\end{abstract}

\end{frontmatter}


\section{Nucleon electromagnetic form factors}
\label{sec2}

Electromagnetic form factors (FFs) of the nucleon
are the standard source of information
on the nucleon structure and as such have been studied extensively; for 
recent reviews see {\it e.g.}  
Refs.~\cite{HydeWright:2004gh,Arrington:2006zm,Perdrisat:2006hj}. 
The most recent generation of electron accelerators, which combine 
high current with high polarization electron beams, at MIT-Bates, the Mainz 
Microtron (MAMI), and the Continuous Electron Beam Accelerator 
Facility (CEBAF) of the Jefferson Lab (JLab), 
have made it possible to investigate the internal structure 
of the nucleon with unprecedented precision. 
In particular, the new generation of polarization experiments 
that make use of the target- and recoil-polarization techniques 
have allowed to chart the proton and neutron electromagnetic (e.m.) FFs 
very precisely. 

\begin{figure}[h]
\begin{center}
{
\includegraphics[width=5cm]{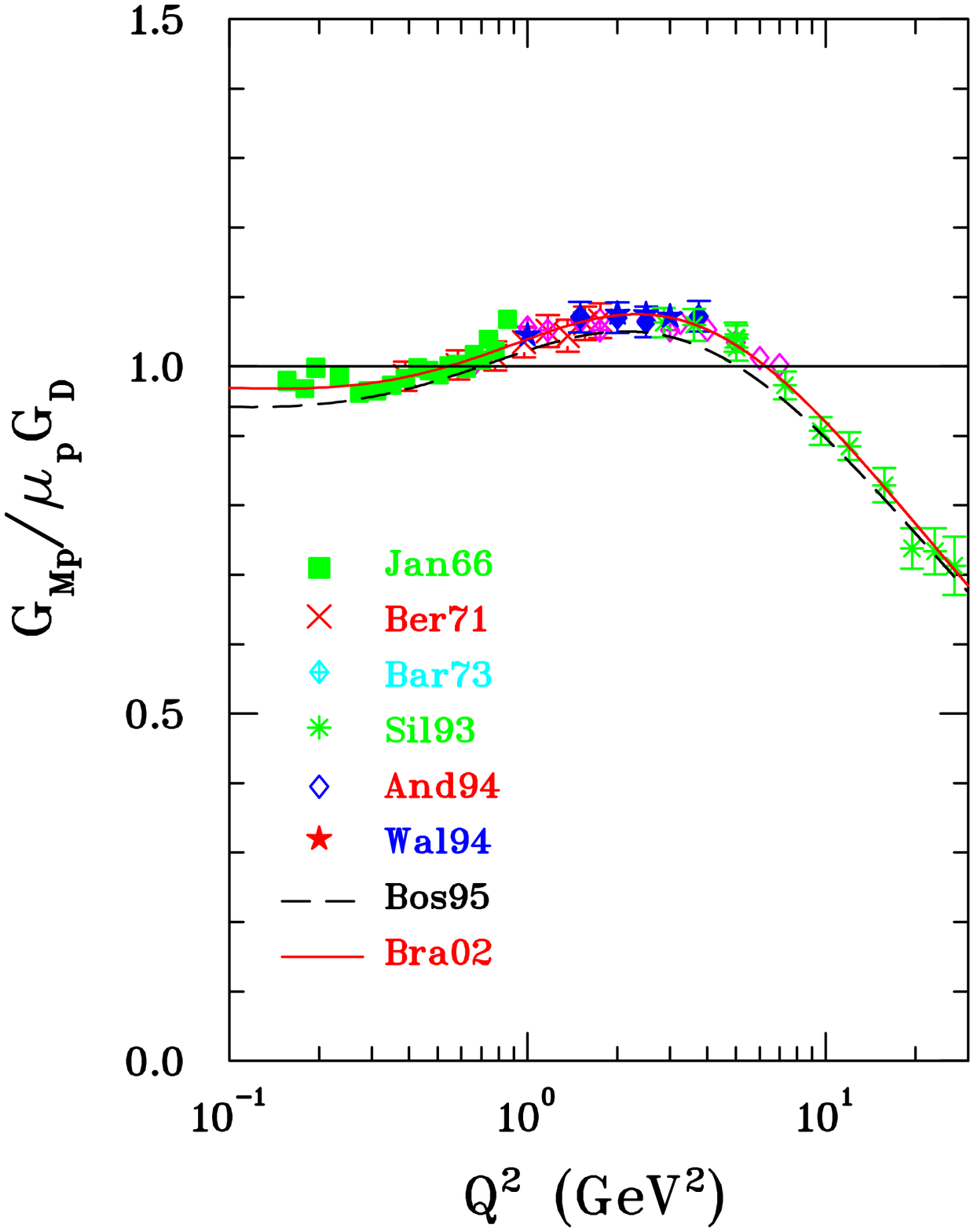} 
\hspace{1cm}
\includegraphics[width=5cm]{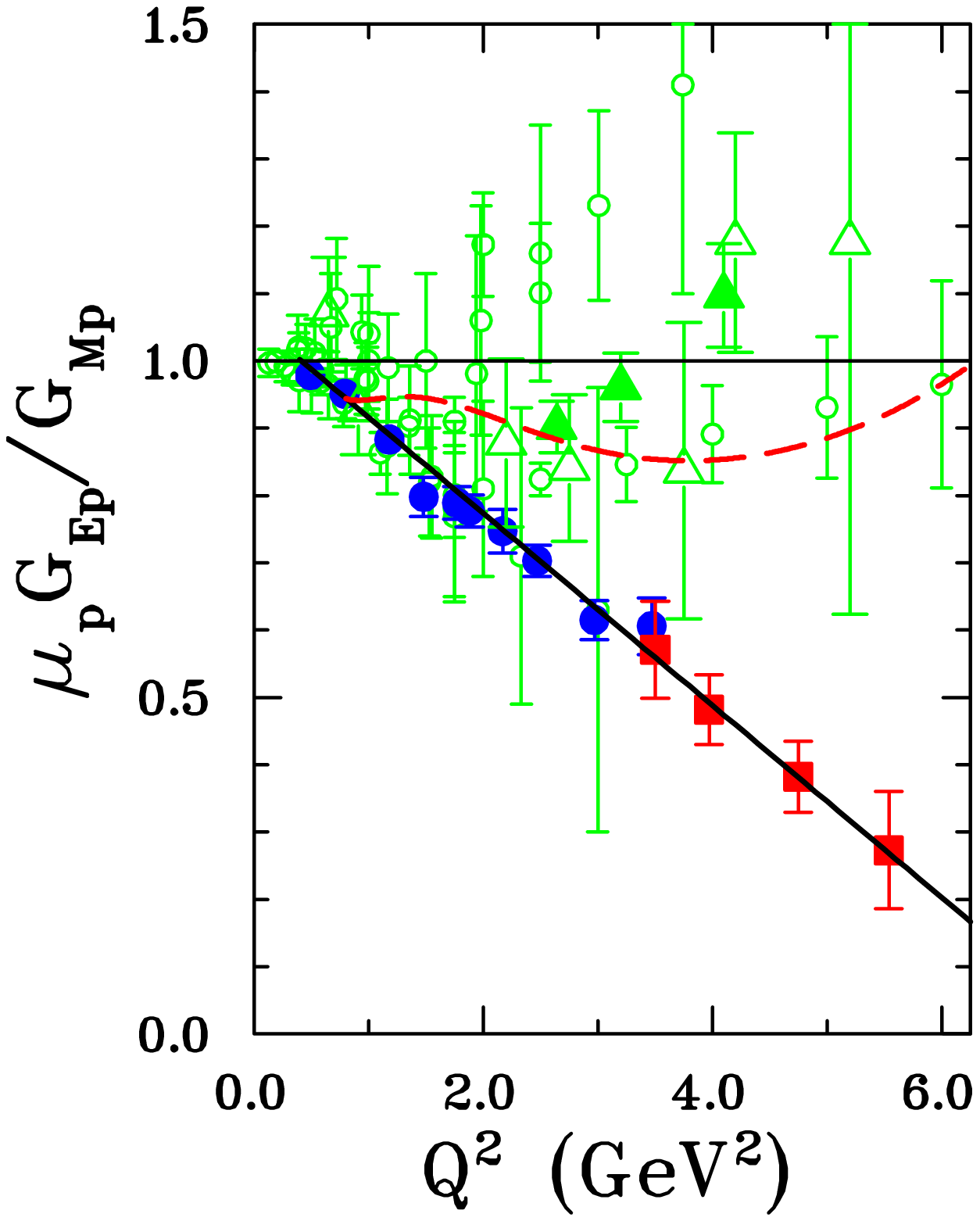}
}
\end{center}
\caption{\small{Left panel : world data base for $G_{M{p}}$ obtained by the 
Rosenbluth method; the references can be found in~\cite{Perdrisat:2006hj}. 
Right panel : comparison of $\mu_p G_{Ep}/G_{Mp}$ from the 
two JLab polarization data 
\cite{Gayou02,Punjabi:2005wq} (solid circles and squares), and 
Rosenbluth separation results (symbols as referred to in 
Ref.~\cite{Perdrisat:2006hj}. 
Dashed curve is a re-fit of Rosenbluth data~\cite{arring03}; 
solid curve is a fit to the JLab polarization data.}}
\label{fig:formf_p}
\end{figure}

The status of proton electric ($G_{Ep}$) and magnetic ($G_{Mp}$)
FF measurements is shown in Fig.~\ref{fig:formf_p} (for more details 
and references, see~\cite{Perdrisat:2006hj}). 
The proton magnetic FF has been measured up to a momentum transfer $Q^2$ of 
around 30~GeV$^2$. 
The deviation of the proton magnetic FF from the standard 
dipole form $G_D = 1/(1 + Q^2 / 0.71)^2$ has been measured precisely,
showing a dip structure at low momentum transfers 
(around $Q^2 \simeq 0.25$~GeV$^2$)  and a scaling behavior 
at very large values of $Q^2$.  
The recent and unexpected results from JLab of using the polarization transfer 
technique to measure the proton electric over magnetic FF ratio, 
$G_{Ep}/ G_{Mp}$~\cite{jones,Gayou02,Punjabi:2005wq}, 
has been the revelation that the FFs obtained using
the polarization and Rosenbluth cross section separation methods, 
were incompatible with each other, starting around $Q^2 = 3$~GeV$^2$ 
(see right panel on Fig.~\ref{fig:formf_p}). 
The FFs obtained from cross section data had 
suggested that $G_{Ep}\sim G_{Mp}/\mu_p$, where $\mu_p$ is the proton magnetic 
moment; the results obtained from recoil polarization data 
clearly show that the ratio 
$G_{Ep}/G_{Mp}$ decreases linearly with increasing $Q^2$. 
The recoil polarization measurement for this ratio has been performed 
up to $Q^2 = 5.6$~GeV$^2$, and a new JLab 
experiment is extending this measurement 
(at the time of writing) up to $Q^2 \simeq 8.5$~GeV$^2$.  
The numerous attempts 
to explain the difference between both experimental techniques 
in terms of radiative corrections which affect 
the results of the Rosenbluth separation method very significantly, 
but polarization 
results only minimally, have led to the previously neglected calculation 
of two hard photon exchange with both photons sharing the momentum transfer, 
as discussed below.    

\begin{figure}[h]
\begin{center}
{
\includegraphics[width=5cm]{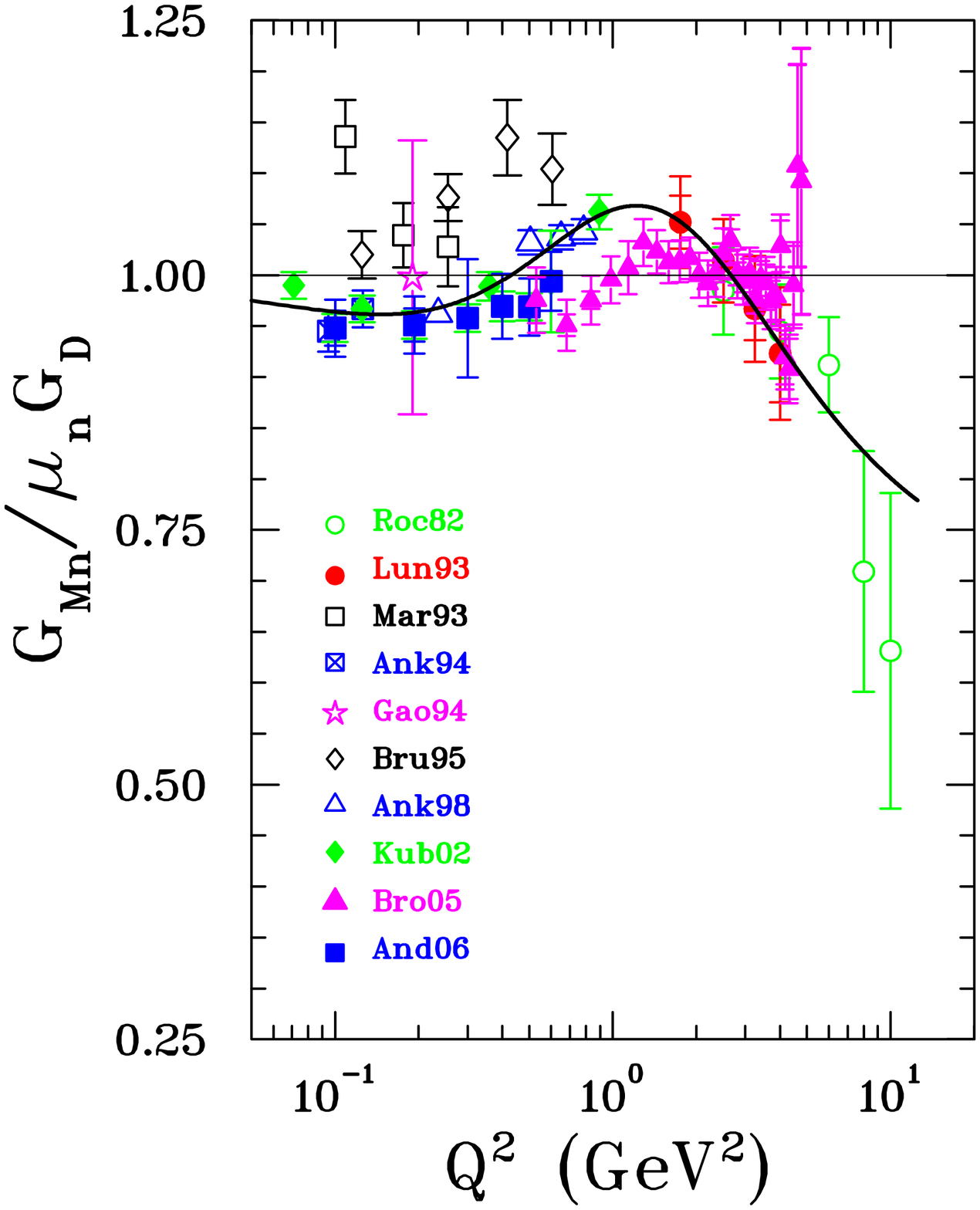} 
\hspace{1cm}
\includegraphics[width=5cm]{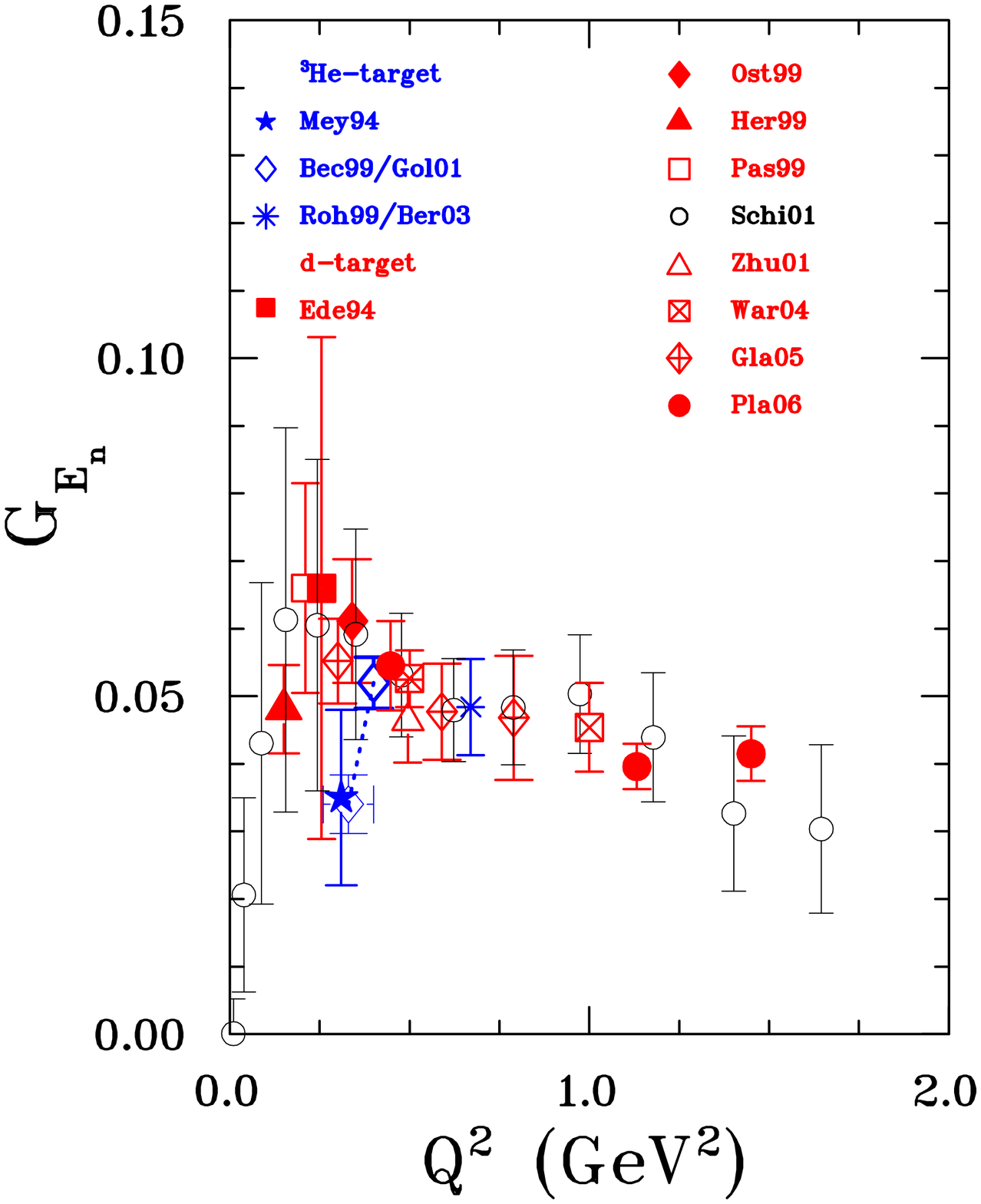}
}
\end{center}
\caption{\small{Left panel : 
world data base for $G_{Mn}$, from cross section and polarization 
measurements.
Shown as a solid curve is the polynomial fit by Kelly \cite{kelly04}. 
Right panel : world data for $G_{En}$ from beam asymmetry with polarized
$D_2$, and $^3$He, and recoil polarization with $D_2$. For references see 
~\cite{Perdrisat:2006hj}.}}
\label{fig:formf_n}
\end{figure}

The use of the polarization technique has also resulted in a 
constant progress in the 
measurement of $G_{En}$, which is intrinsically more difficult to obtain 
because of the smallness of this FF, 
due to the overall zero charge of the neutron. 
Recent times have seen the maximum $Q^2$ for which we have polarization FFs  
grow to 1.5 GeV$^2$, with new data 
obtained and under analysis up to 3.4 GeV$^2$, and several experiments 
planned or proposed to significantly higher $Q^2$ values. 
Important progress has been made for $G_{Mn}$ too, 
with new data with much improved error bars up 
to 4.8 GeV$^2$, see Fig.~\ref{fig:formf_n}. 

When comparing the precision data for the nucleon e.m. FFs in 
Figs.~\ref{fig:formf_p} and \ref{fig:formf_n}, 
Friedrich and Walcher ~\cite{Friedrich:2003iz} made the striking 
observation that all four FFs display ``bump structures'' 
around $Q^2 \simeq 0.25$~GeV$^2$. 
They interpret this structure as a signature of the pion cloud. 
For a comparison with new data of the BLAST Coll. at low $Q^2$  
for both proton and neutron, see~\cite{kohl_inpc}.  

Given the large amount of precise data on FFs it is of 
interest to exhibit directly the spatial information which 
results from these data. When viewing the nucleon 
from a light front moving towards the nucleon, a model independent 
2-dimensional mapping of the quark charge density in the nucleon can be 
achieved in the transverse plane (perpendicular to the direction of motion). 
Using only the empirical information on the nucleon 
e.m. FFs, these transverse charge densities 
have been extracted recently for an unpolarized nucleon~\cite{Miller:2007uy}, 
as well as for a a transversely polarized nucleon~\cite{Carlson:2007xd}. 

\begin{figure}[h]
\begin{center}
{
\includegraphics[width =5.75cm]{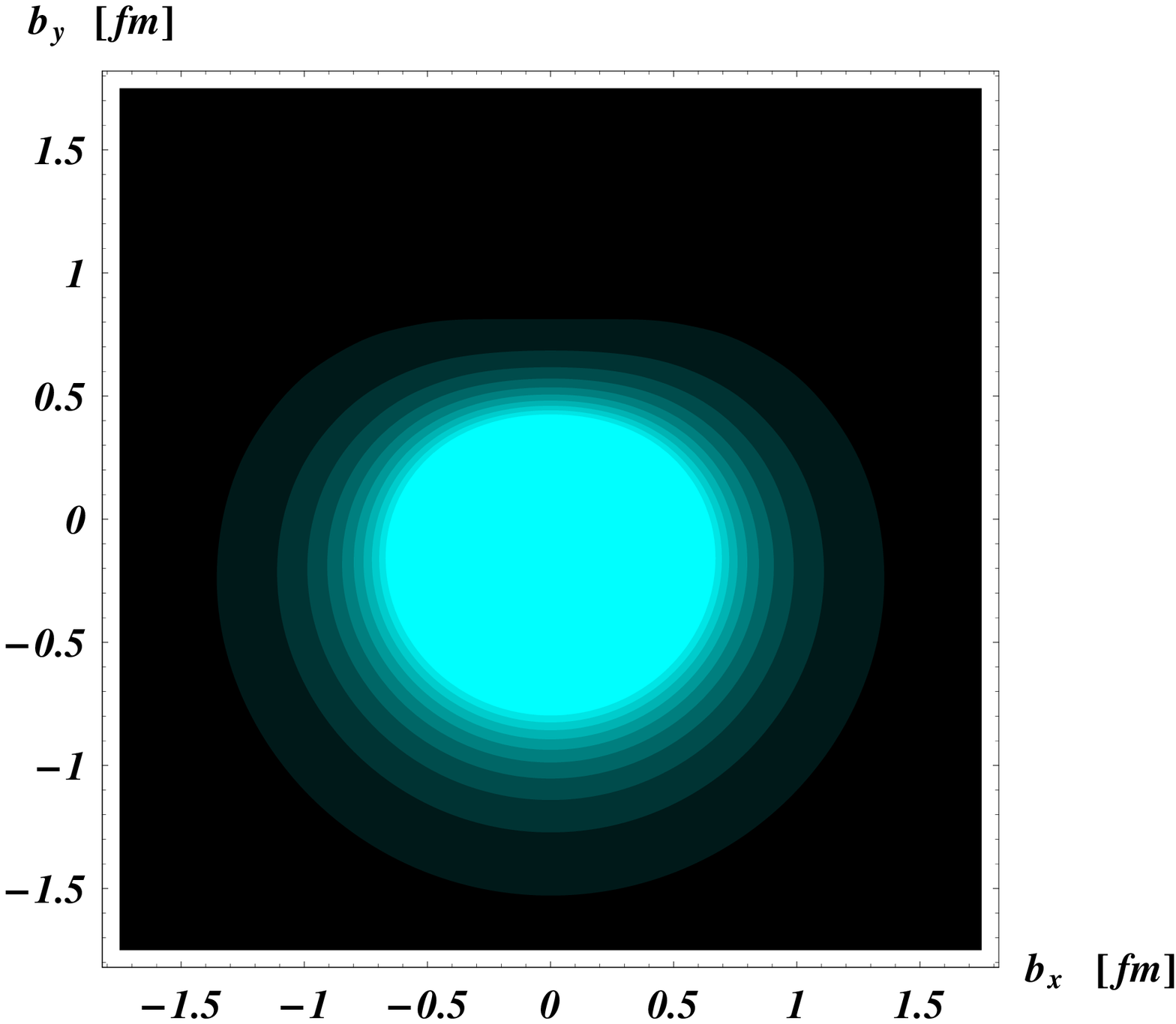}
\hspace{.5cm}
\includegraphics[width =5.75cm]{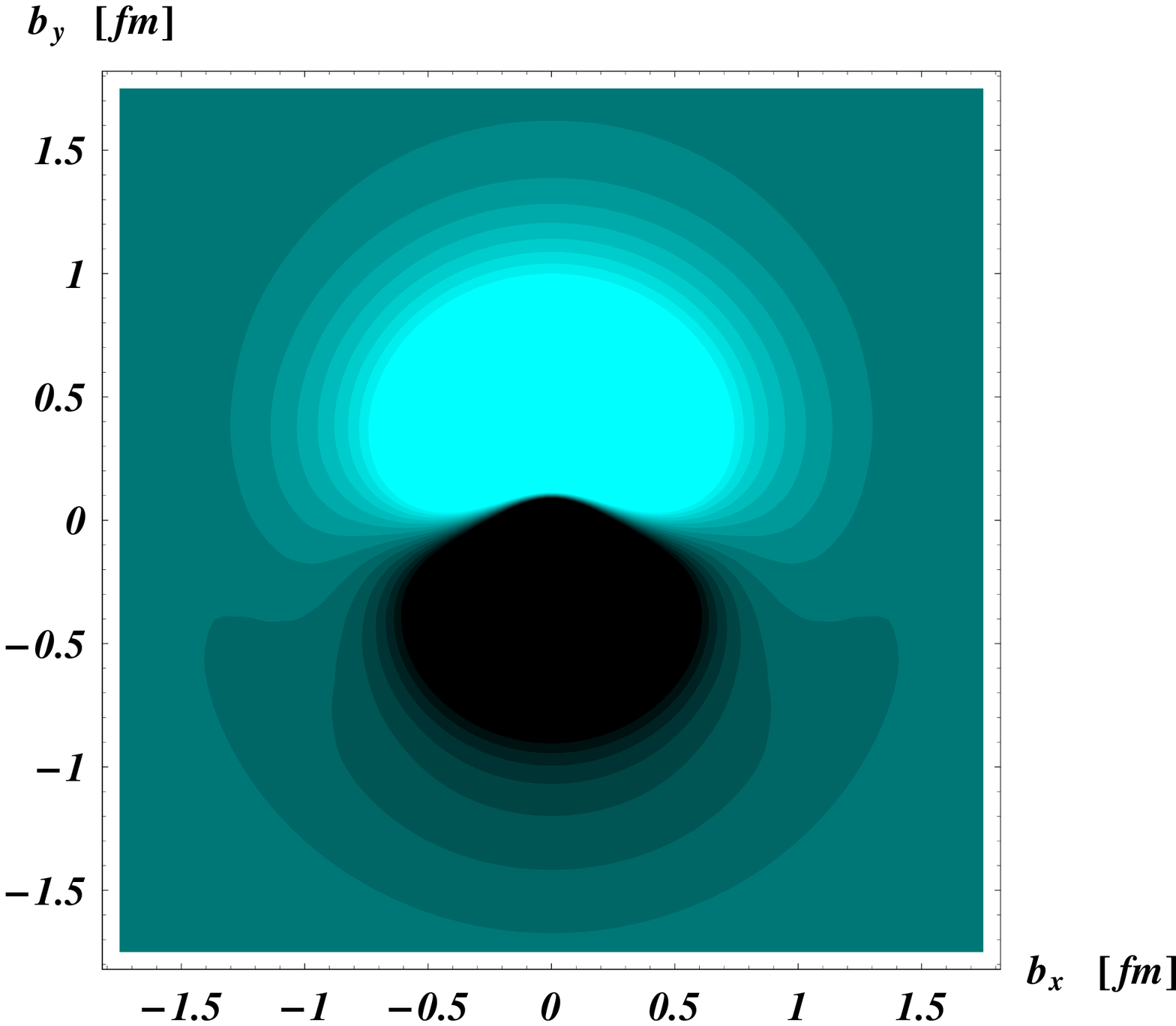}
}
\end{center}
\begin{center}
{
\includegraphics[width =6cm]{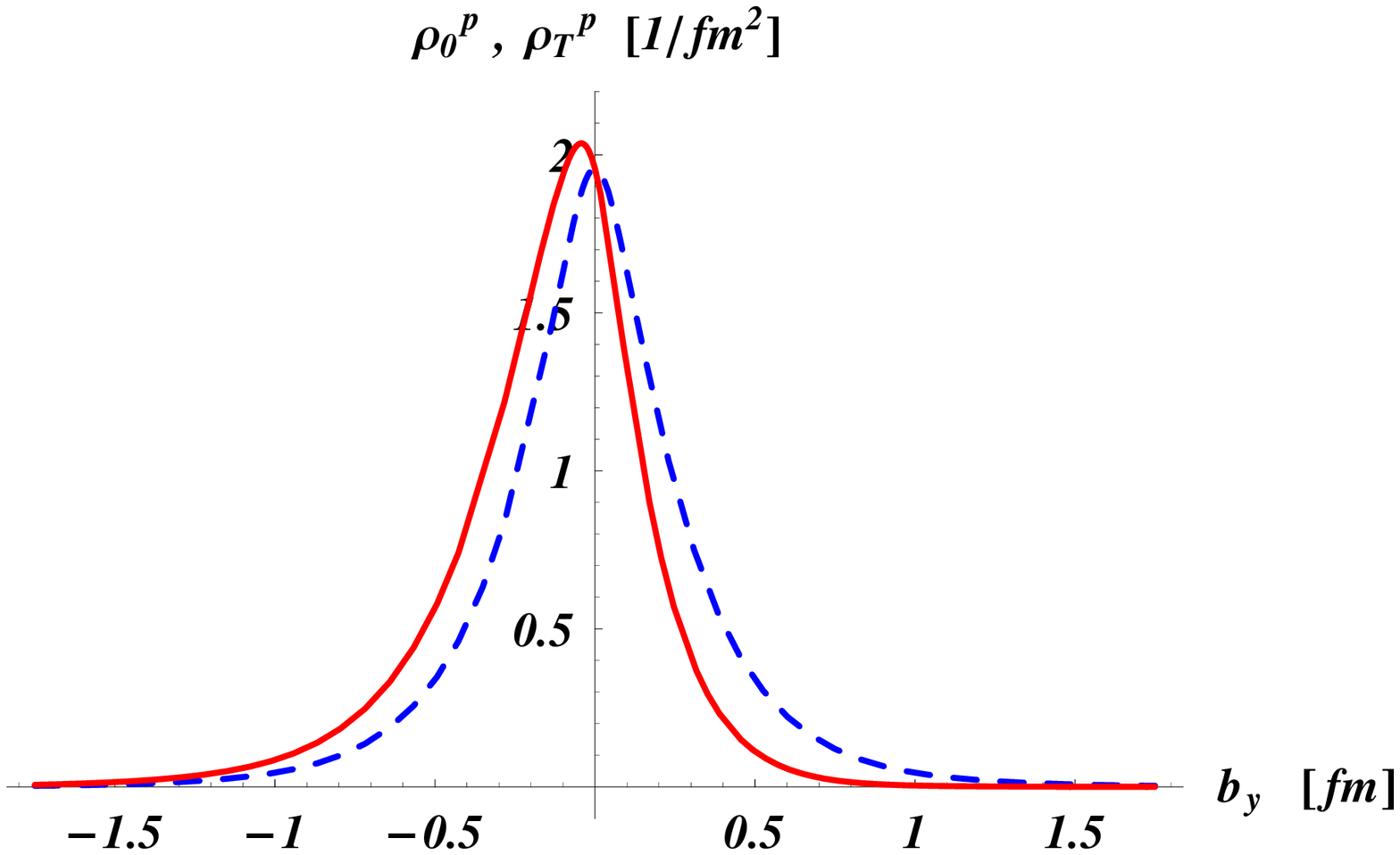}
\hspace{.5cm}
\includegraphics[width =6cm]{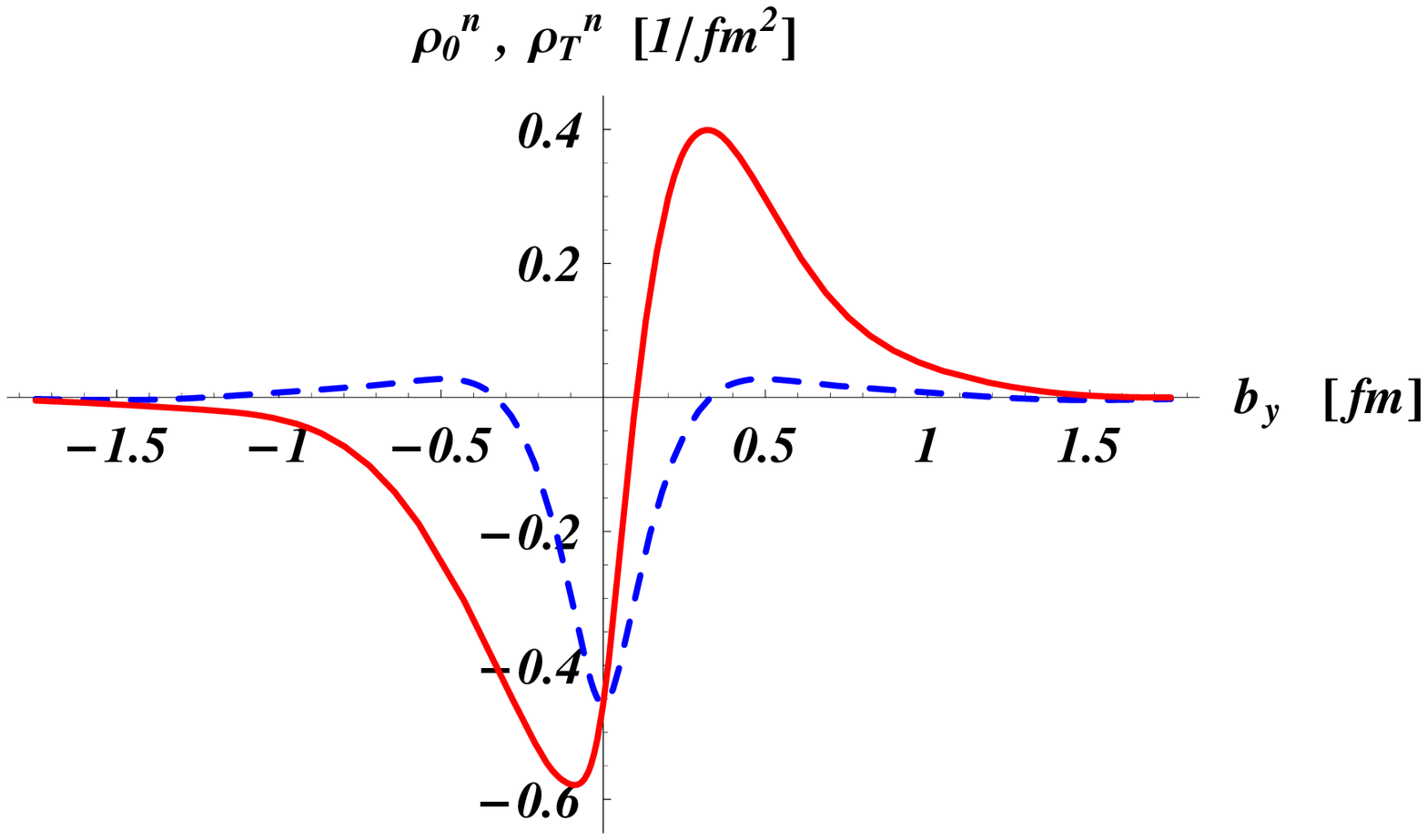}
}
\end{center}
\caption{Quark transverse charge densities in the {\it proton} (left panels) 
and {\it neutron} (right panels). 
The upper panel shows the density in the transverse plane for a 
nucleon polarized along the $x$-axis. The light (dark) regions correspond with
largest (smallest) values of the density. 
The lower panel compares the density along the $y$-axis 
for an unpolarized nucleon (dashed curve), 
and for a nucleon polarized along the $x$-axis (solid curve). 
For the proton (neutron) e.m. FFs, the empirical parameterization of 
Arrington {\it et al.}~\cite{Arrington:2007ux} 
(Bradford {\it et al.}~\cite{Bradford:2006yz}) is used. 
Figure from~\cite{Carlson:2007xd}.}
\label{fig:densities}
\end{figure}

In Fig.~\ref{fig:densities} these transverse charge densities are shown 
for proton and neutron, both for the unpolarized case and for a 
nucleon polarized along the $x$-axis (denoting the direction of the fast
moving frame by the $z$-axis).  
One notices from Fig.~\ref{fig:densities} that 
polarizing the proton along the $x$-axis 
leads to an induced electric dipole field which 
corresponds with an electric dipole moment along the 
negative $y$-axis equal to the value of the 
anomalous magnetic moment, {\it i.e.} $\kappa_N$ (in units $e/2 M_N$) as  
noticed in~\cite{Burkardt:2000za}. 
One sees that the corresponding neutron's transverse 
charge density gets displaced significantly due to the large 
(negative) value of the neutron anomalous magnetic moment,  
$\kappa_n = -1.91$, which yields an induced electric dipole moment along the 
positive $y$-axis.  

To calculate nucleon e.m. FFs from first principles, lattice QCD simulations 
have seen an important progress in recent years. 
State-of-the-art lattice calculations for nucleon structure studies  
use lattice spacings $a\stackrel{<}{\sim}0.1$~fm and
lattice sizes $L\sim 3$~fm and reach pion mass values down to about 350~MeV.
To connect those results with the physical world requires an extrapolation 
down to the physical quark mass $m_q$ 
(with $m_q$ proportional to $m_\pi^2$). 
It is only very recently that pion mass values 
below $350$~MeV~\cite{schaefer_inpc} have been reached. 
This continuous effort is important to eliminate
one source of systematic error associated with the extrapolation to 
the light quark masses. 

The lattice calculations for the (space-like) nucleon e.m. FFs require 
the evaluation of three-point functions, which 
involve two topologically different contributions.
In the connected diagram contribution, 
the photon couples to one of the quarks connected 
to either the initial or final nucleon. 
The disconnected diagram, which involves a coupling to a 
$q \bar q$ loop, requires a numerically 
more intensive calculation, is at present neglected in most lattice studies. 
When taking the difference between proton and neutron e.m. FFs, i.e. 
for the isovector combination, the disconnected contribution drops out. 
Therefore, the calculations in which the disconnected 
diagram is neglected are applicable only to the isovector e.m. FFs. 

\begin{figure}
\hspace{.5cm}
\epsfxsize=6cm \epsffile{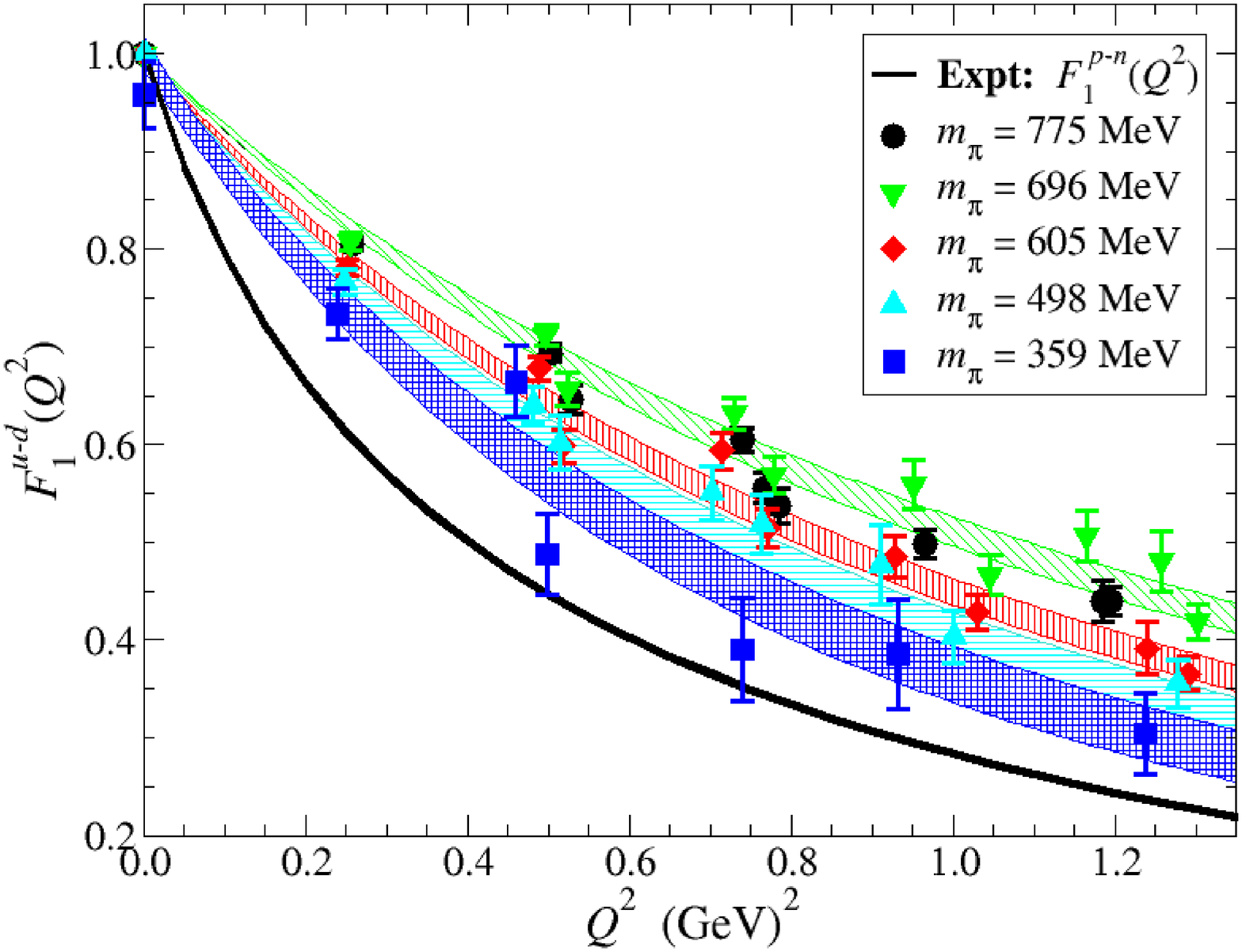} 
\hspace{.5cm}
\epsfxsize=6.25cm \epsffile{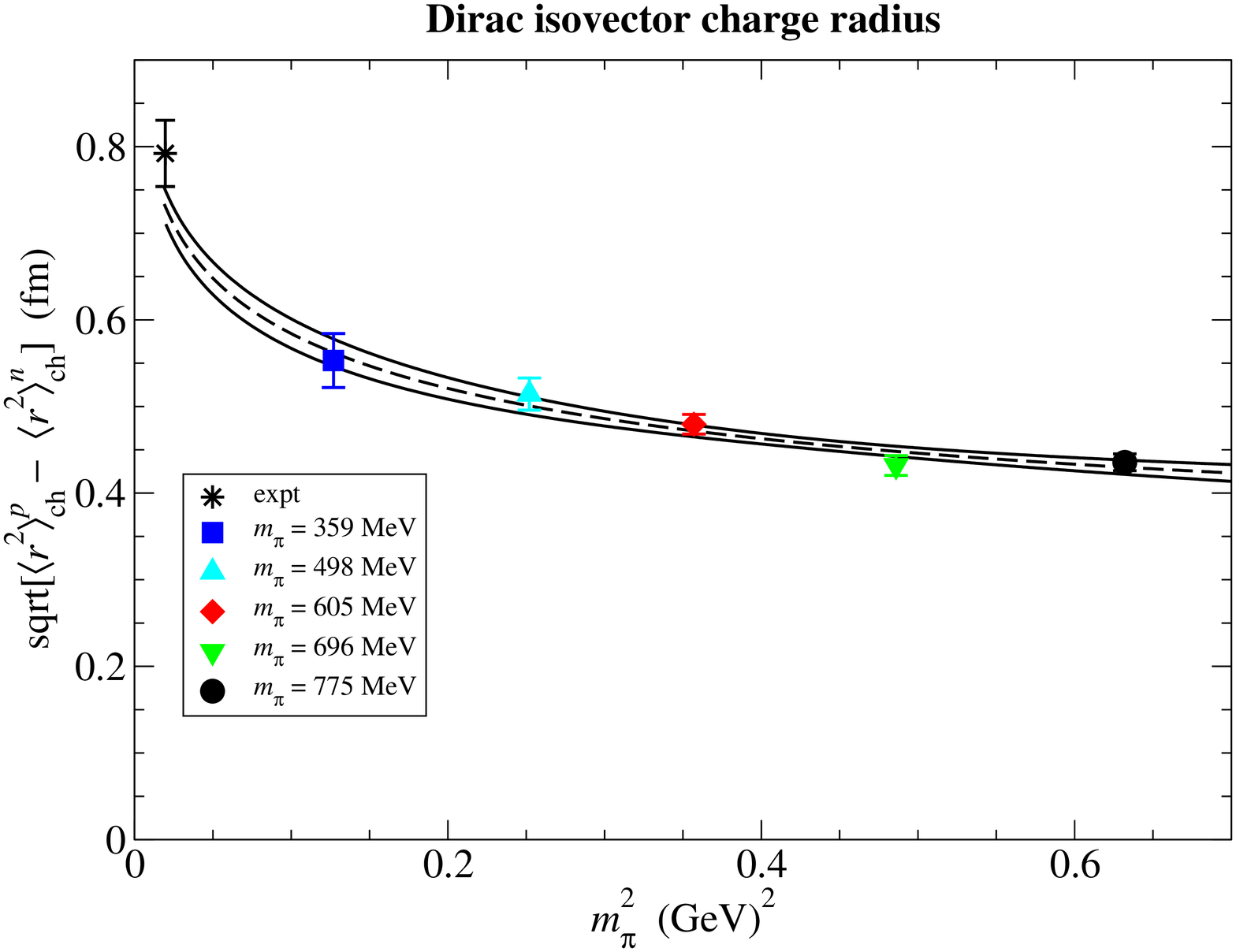}
\caption{\small 
Left panel : Lattice QCD results (from the LHPC Coll.~\cite{Edwards:2006qx})   
for the nucleon isovector FF $F_1^V$. The unquenched results 
using  a hybrid action of domain wall valence quarks and 
2+1 flavor staggered sea quarks are shown for   
different values of $m_\pi$ and are compared with experiment 
(solid curve, using the parameterization of~\cite{kelly04}).   
Right panel : Chiral extrapolation of the nucleon isovector Dirac radius 
$\langle r^2 \rangle_1^V$.   
The unquenched results are from the LHPC Coll.~\cite{Edwards:2006qx}. 
The experimental value is shown by the star. The curves are fits using the 
chiral extrapolation formula Eq.~(\ref{eq:rms-extrap}). 
} 
\label{fig:f1f2_lhpc}
\end{figure}

Fig.~\ref{fig:f1f2_lhpc} shows the unquenched lattice QCD 
results from the LHPC Coll. for the nucleon e.m. FFs, 
performed for one lattice spacing of 
$a \simeq 0.125$~fm, and for pion mass values in the 
range $m_\pi = 360 - 775$~MeV. It is seen that this action yields a 
noticeable dependence on $m_\pi$ for the Dirac isovector FF $F_1^V$. 
The $Q^2$ dependence of $F_1^V$ at the 
smallest $m_\pi$ value of around 360~MeV is found 
to be in qualitative agreement with the data. 
Evidently, it will be very worthwhile to corroborate the results at the
lowest pion masses and improve their statistics in future calculations. 
For lattice calculations of the isovector FFs using 
Wilson fermions, from the QCDSF Coll., 
see~\cite{schaefer_inpc}; and for calculations using  
dynamical domain wall fermions, from the RBC and UKQCD Coll., 
see \cite{Ohta_inpc}. 

Present lattice calculations are possible for larger than physical  
quark masses, and therefore necessitate an extrapolation 
procedure in order to make contact with experiment. 
The extrapolation in $m_q$ is not straightforward,
because the non-analytic dependencies, such as $\sqrt{m_q}$
and $\ln m_q$, become important as one approaches the
small physical value of $m_q$. As an example, the Dirac charge radius, 
$\langle r^2 \rangle_1^V$ varies as $\ln m_\pi^2$ when approaching the 
chiral limit. To extend the range of validity in $m_\pi$ of such a 
prediction, a one may try in this spirit a modification 
of the chiral perturbation theory formula as~\cite{Dunne:2001ip}~:
\begin{eqnarray}
\langle r^2 \rangle_1^V  
= a_0
-\frac{1+5g_A^2}{(4 \pi f_\pi)^2}\ln \left(\frac{m^2_\pi}{m_\pi^2 + 
\Lambda^2}\right),
\label{eq:rms-extrap}
\end{eqnarray}
where $\Lambda$ is a phenomenological cut-off which reflects the finite size 
of the nucleon, with $a_0$ a low-energy constant. The other parameters in this
expression are fixed and well known. 
Such a fit (using $\Lambda \sim 500$~MeV) for the isovector Dirac radius 
is shown in Fig.~\ref{fig:f1f2_lhpc} and compared with the most recent 
unquenched lattice results using the hybrid action 
(domain wall valence quarks on top of a 2+1 flavor staggered sea) 
of the LHPC Coll. 
One firstly sees, that these lattice results do show appreciable $m_\pi^2$ 
variation over the pion mass range $m_\pi = 360 - 775$~MeV and provide a 
first clear hint of the logarithmic $m_\pi$ divergence. 
As the pion mass approaches the physical value, the calculated nucleon 
size increases and approaches the correct value.

\begin{figure}[h]
\leftline{\includegraphics[width=6cm]{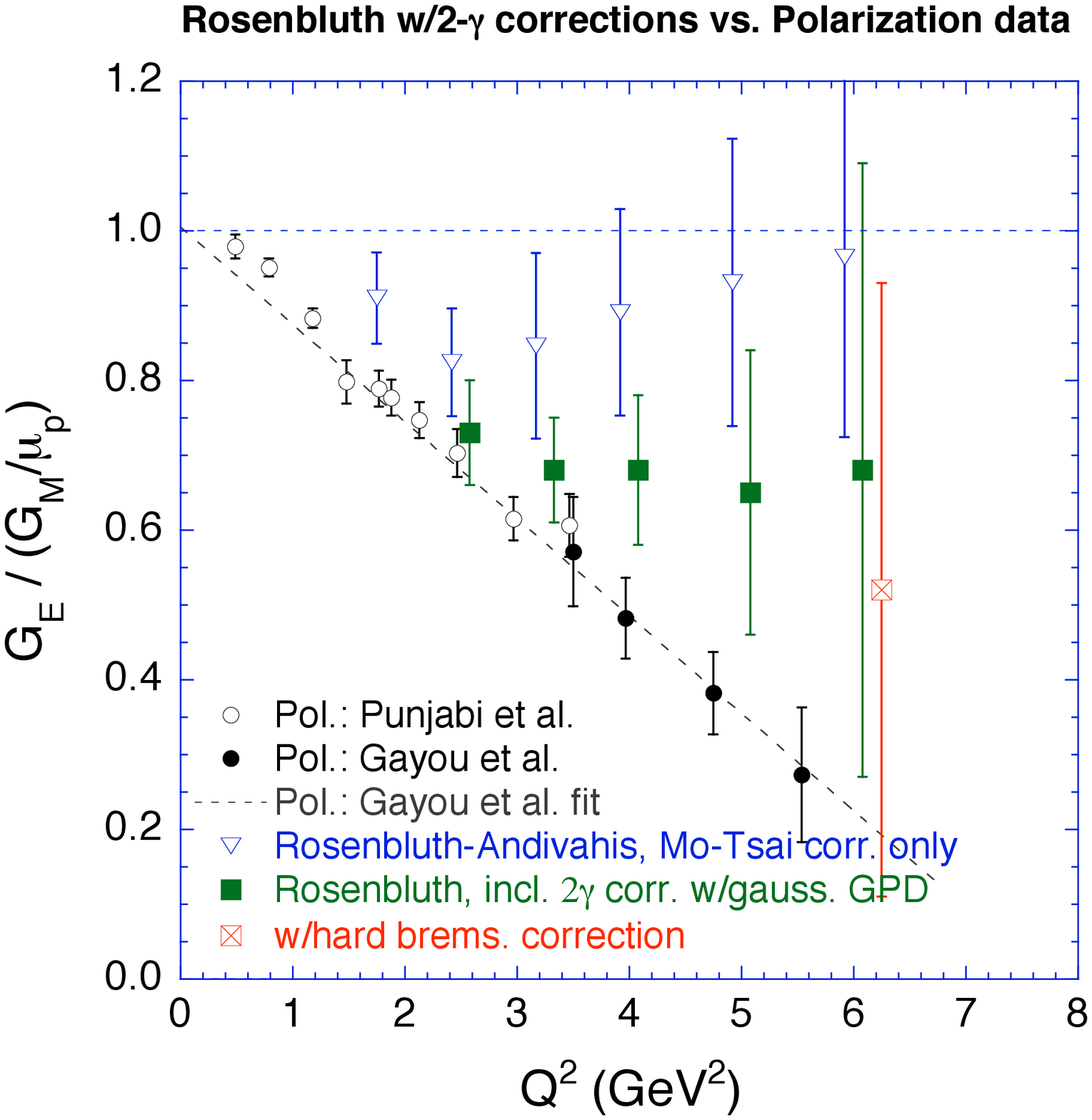}}
\vspace{-6cm}
\rightline{\includegraphics[width=5.5cm,angle=-90]{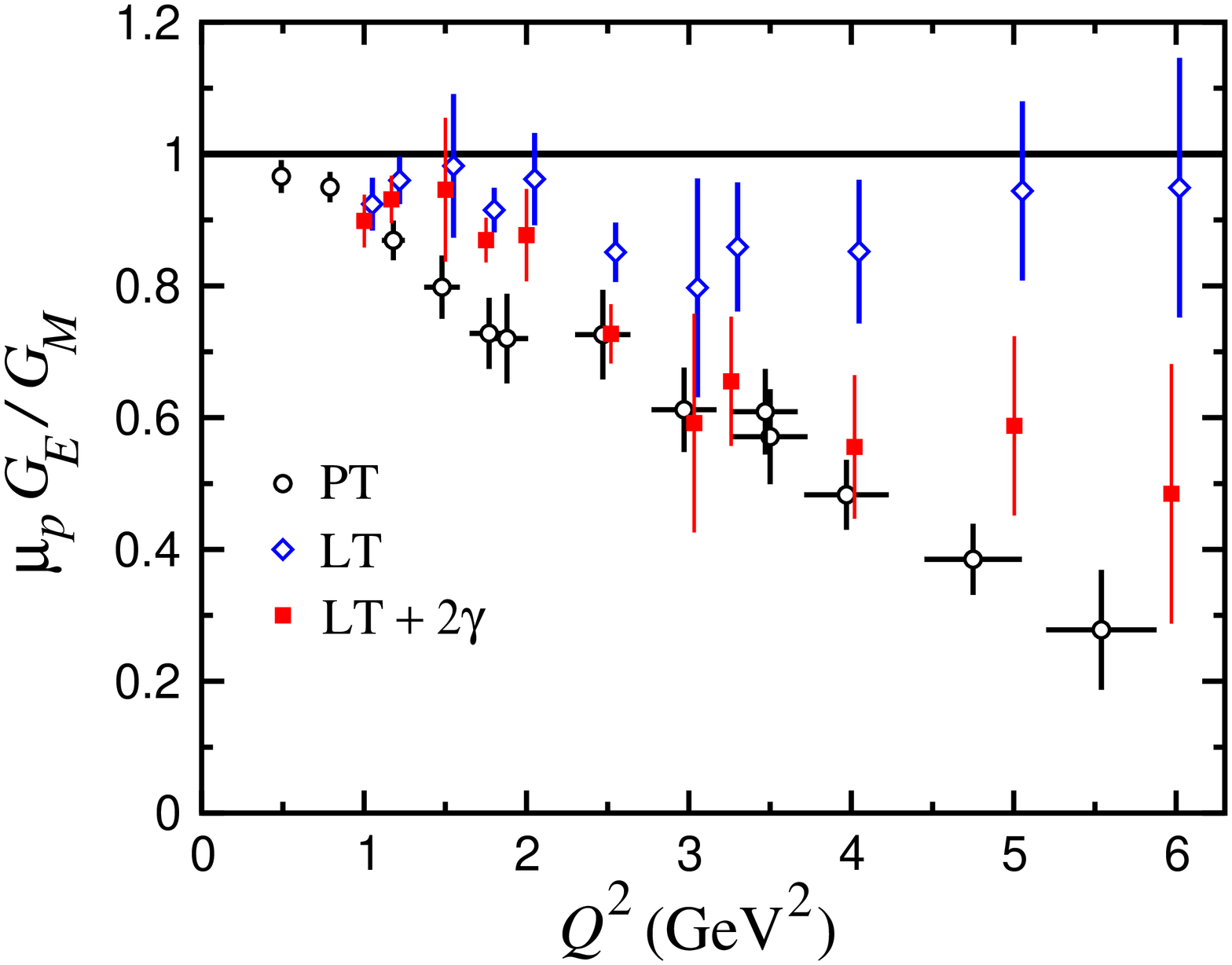}}
\caption{Determinations of the proton $G_E/G_M$ ratio 
with $2\gamma$ corrections calculated 
within the partonic framework (left panel)~\cite{YCC04,ABCCV05} or using a 
single proton as the hadronic intermediate state 
(right panel)~\cite{BMT03}. 
Left panel : polarization data from~\cite{Gayou02,Punjabi:2005wq}, 
Rosenbluth data from~\cite{Slac94}, 
which include only the well-known Mo-Tsai corrections. 
The Rosenbluth $G_E/G_M$ include the $2 \gamma$ corrections, 
and for one point also a hard bremsstrahlung correction, 
still using Andivahis {\it et al.}~data.  
Right panel : 
``PT'' is $G_E/G_M$ obtained from the polarization transfer experiments;
``LT'' is $G_E/G_M$ obtained from a Rosenbluth experiment using 
only the (Mo-Tsai) radiative corrections, and 
``LT+$2\gamma$'' includes the extra $2\gamma$ corrections.  
}
\label{newcomparison}
\end{figure}

At larger momentum transfers, 
the striking difference between the unpolarized (Rosenbluth) and 
polarization transfer measurements of the proton $G_{Ep} / G_{Mp}$ FF 
ratio has triggered a renewed interest in the field of 
two-photon exchange in $eN$ scattering experiments. 
Theoretical calculations both within a hadronic and partonic framework 
made it very likely that hard two-photon exchange corrections 
are the main culprit in the difference between both experimental 
techniques~\cite{GV03}. 
Despite the long history of two-photon exchange corrections, 
see Ref.~\cite{Carlson:2007sp} for a review, 
it is interesting to note that concepts developed over the past decade, such
as generalized parton distributions which describe two-photon processes 
with one or two large photon virtualities, enter when quantifying 
two-photon exchange corrections at larger $Q^2$. 
The model-independent finding is that the hard two-photon corrections  
hardly affect polarization transfer results, but they 
do correct the slope of the Rosenbluth plots at larger $Q^2$ in an important
way, towards reconciling both experimental techniques. 

As an example, we show in Fig.~\ref{newcomparison} (left panel)  
the $2$-$\gamma$ exchange correction 
on the extracted $G_{Ep} / G_{Mp}$ vs. $Q^2$ within a partonic 
framework~\cite{YCC04,ABCCV05}.  
One set of data points, falling linearly with $Q^2$, 
is from the polarization experiments.  Another set of data points, 
roughly constant in $Q^2$ and plotted with inverted triangles, 
is from Rosenbluth data analyzed using only radiative corrections.  
The solid squares show the best fit $G_{E} / G_{M}$ from~\cite{Slac94} 
data when analyzed including the hard $2$-$\gamma$ corrections.  
One sees that for $Q^2$ in the 2--3 GeV$^2$ range, 
the $G_{Ep}/G_{Mp}$ extracted using the Rosenbluth method including the 
$2$-$\gamma$ corrections agree well with the polarization transfer results.  
At higher $Q^2$, there is at least partial reconciliation between the 
two methods.   
The effect, in a calculation with just a proton intermediate state, 
see e.g.~\cite{BMT03}, 
is qualitatively similar and shown on the right panel in 
Fig.~\ref{newcomparison}.  
Several forthcoming high-precision electron scattering experiments 
aim at a precision test of these two-photon exchange effects.

\section{Generalized Parton Distributions}
\label{sec3}

\begin{figure}[t]
\epsfysize=3cm
\epsffile{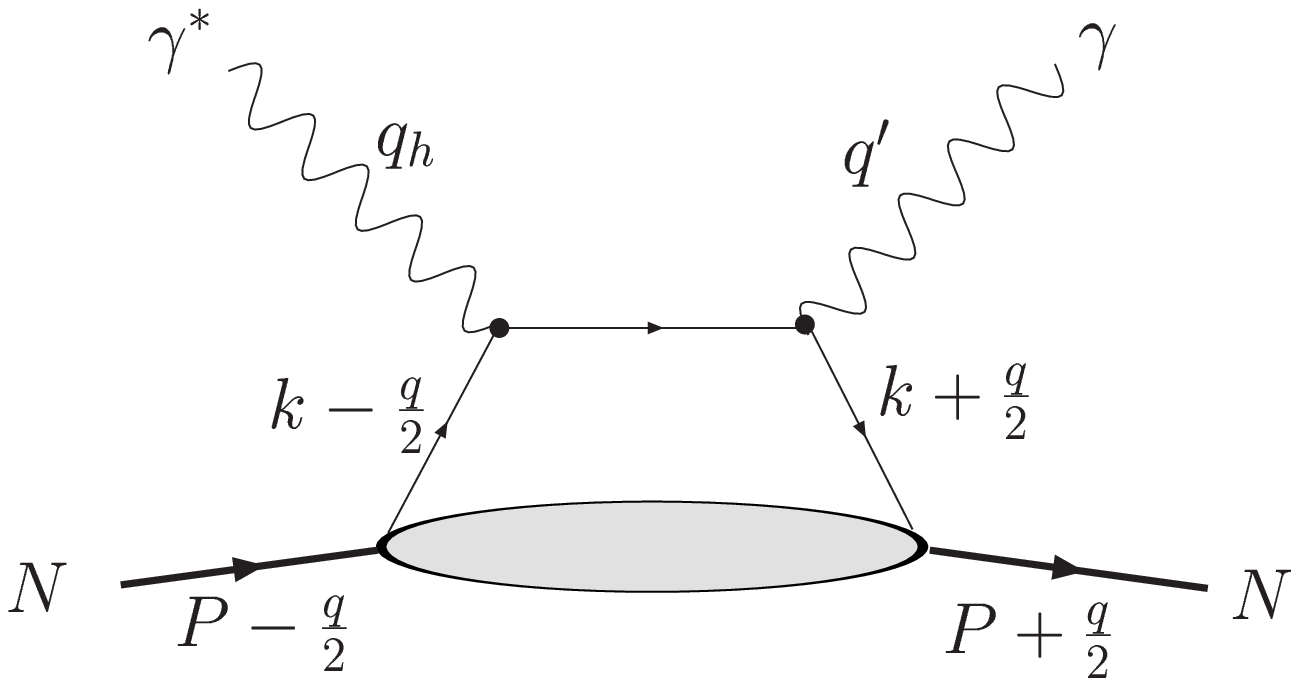} 
\hspace{.5cm}
\epsfysize=4.5cm
\epsffile{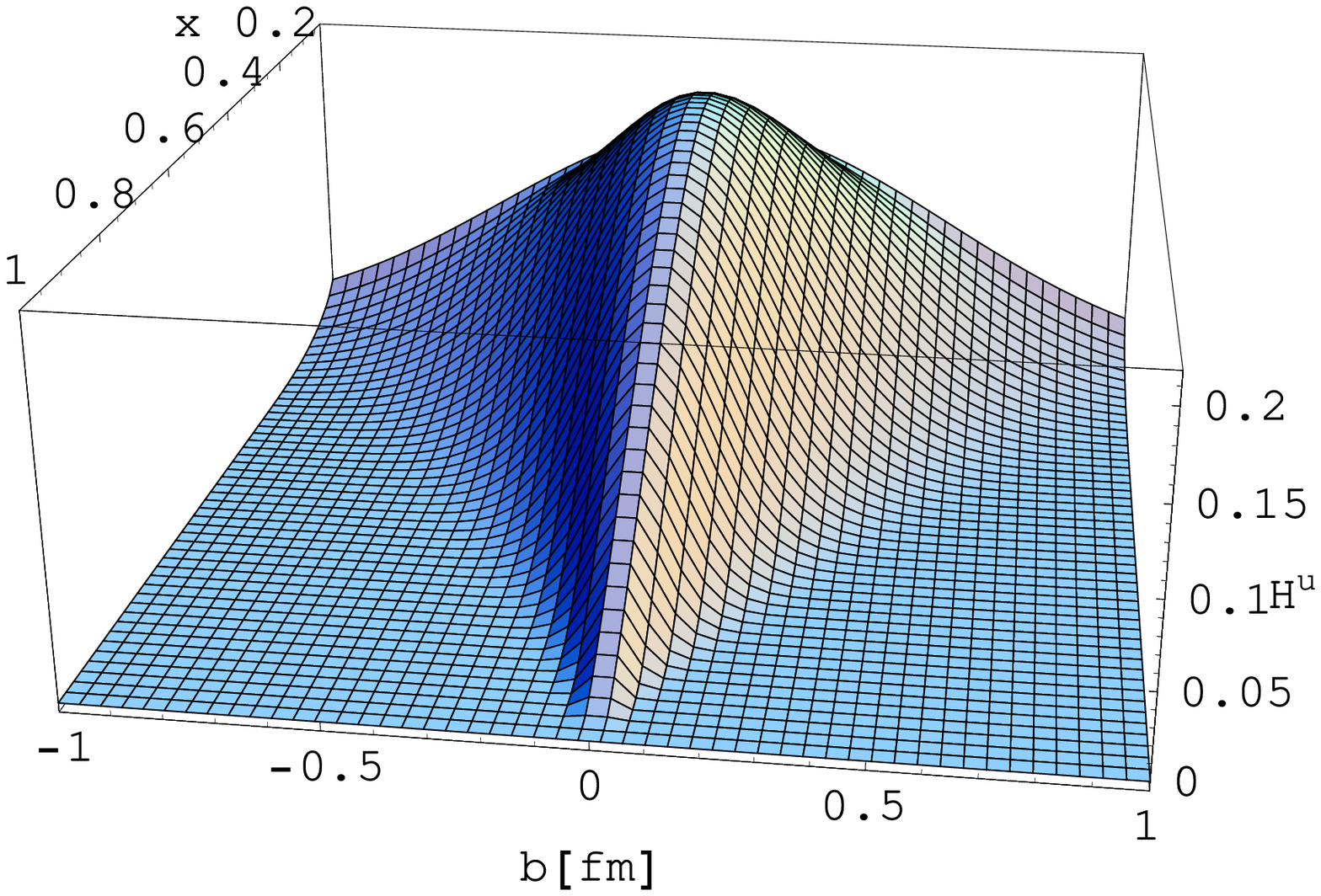} 
\caption{\small 
Left panel : ``handbag'' diagram for the nucleon DVCS process. 
Provided the virtuality of the initial photon (with momentum $q_h$) 
is sufficiently large, the 
QCD factorization theorem allows to express the 
total amplitude as the convolution  
of a Compton process at the quark level and a non-perturbative 
amplitude parameterized in terms of GPDs (lower blob).    
Right panel : the GPD $H^u$ for a valence up-quark in the proton  
as function of the quark momentum fraction $x$ and 
the quark position $b$ in the transverse plane (perpendicular to the 
average direction of the fast moving nucleons). 
The calculation is based on the 3-parameter modified Regge parametrization   
of \cite{guidal}. 
}
\label{fig:gpd}
\end{figure}

The nucleon e.m. FFs discussed above access the quark-gluon structure 
of the nucleon by measuring the matrix 
element of a  well-defined quark-gluon operator (in this case the vector 
operator $\bar q \gamma^\mu q$) over the hadronic state. 
One is however not limited in nature to probes such as photons 
(or $W$, $Z$ bosons for the axial transition). The phenomenon of asymptotic 
freedom of QCD, meaning that at short distances the interactions between 
quarks and gluons become weak, provides us with more sophisticated 
QCD operators to explore the structure of hadrons. Such operators can 
be accessed by selecting a small size configuration of quarks and gluons, 
provided by a hard reaction, such as deep inelastic scattering (DIS), or 
hard exclusive reactions such as deeply virtual Compton scattering (DVCS), 
$\gamma^*(q_h) + N(p) \to \gamma(q^\prime) + N(p^\prime)$, where the 
virtual photon momentum $q_h$ is the hard scale, 
see left panel of Fig.~\ref{fig:gpd}.   
The common important feature of such hard reactions is the possibility
to separate clearly the perturbative and nonperturbative stages of
the interactions~: this is the so-called factorization property. 
The non-perturbative stage of such hard exclusive electroproduction 
processes is described by universal objects, so-called 
generalized parton distributions (GPDs), 
see~\cite{Ji:1998pc,Goeke:2001tz,Diehl:2003ny,Ji:2004gf,Belitsky:2005qn} 
for reviews and references.

The nucleon structure information entering the nucleon DVCS process, 
can be parameterized at leading twist-2 level, in
terms of four (quark chirality conserving) GPDs, depending  
on three variables: the quark longitudinal 
momentum fractions $(x + \xi)$ of intial quark, $(x - \xi)$ of final quark, 
and the momentum transfer $Q^2 = - q^2$ to the nucleon.
The interplay between the $x$ and $Q^2$-dependence of the GPDs 
contains new nucleon structure information beyond the information encoded 
in forward parton distributions depending only on $x$, or FFs 
depending only on $Q^2$. 
A Fourier transform of the $Q^2$-dependence of GPDs accesses the
distributions of parton in the transverse plane~\cite{Burkardt:2000za}.  
For the vector GPD $H^q$ at $\xi = 0$, this Fourier integral 
in transverse momentum $q_\perp$ reads as:
\begin{eqnarray}
H^q(x, {\bf b_\perp}) \,& \equiv &\, 
\int \frac{d^2 {\bf q_\perp}}{(2 \pi)^2} \, 
e^{i {\bf b_\perp \cdot q_\perp}} \;
H^q (x, \xi = 0,- {\bf q_\perp^2}) ,  
\end{eqnarray}
and an analogous definition for the other GPDs. 
These impact parameter GPDs have the physical meaning of measuring the 
probability to find a quark which carries longitudinal 
momentum fraction $x$ at a transverse position 
${\bf b_\perp}$ (relative to the transverse center-of-momentum) in a nucleon, 
see~\cite{Burkardt:2000za}.  
The right panel of Fig.~\ref{fig:gpd} shows the GPD $H^u$ , for an
$u$-quark, within a 3-parameter modified Regge 
GPD parameterization~\cite{guidal}. 

Besides providing a tomographic view of the nucleon, GPDs 
also allow an access to the angular momentum of quark ( and gluons ) in
the nucleon~\cite{Ji:1996ek}, for a comparison with recent lattice 
calculations see~\cite{schaefer_inpc}.  

On the experimental side, 
the first round of experiments dedicated to measure 
hard exclusive processes such as deeply virtual Compton scattering have 
been completed at HERMES, HERA, and JLab@6GeV, indicating the dominance 
of the twist-2 mechanism. For a discussion of some of these 
results, see~\cite{bertin_inpc}. 
Furthermore, accessing the nucleon GPDs is a major project 
for the planned JLab 12 GeV upgrade.

\section{Nucleon excitation spectrum}
\label{sec4}

The understanding of the nucleon excitation spectrum is tightly linked to 
its structure. The first baryon excited state, the $\Delta(1232)$ resonance 
has been charted in particular detail in recent years, see 
\cite{Pascalutsa:2006up} for a recent review and references. 
At low $Q^2$, the measurement of $\gamma N \Delta$ FFs 
have allowed to unambiguously map out 
the small $E2$ and $C2$ amplitudes revealing $d$-wave components 
in the nucleon and / or $\Delta$ wave functions. 
In parallel, recent years have also seen the development of a chiral effective 
field theory ($\chi$EFT) 
with two distinct light mass scales : the pion mass and the 
$\Delta - N$ mass 
difference~\cite{JeM91a,HHK97,Pascalutsa:2002pi,Pascalutsa:2005}.

\begin{figure}[h]
\leftline{\includegraphics[height=8.5cm]{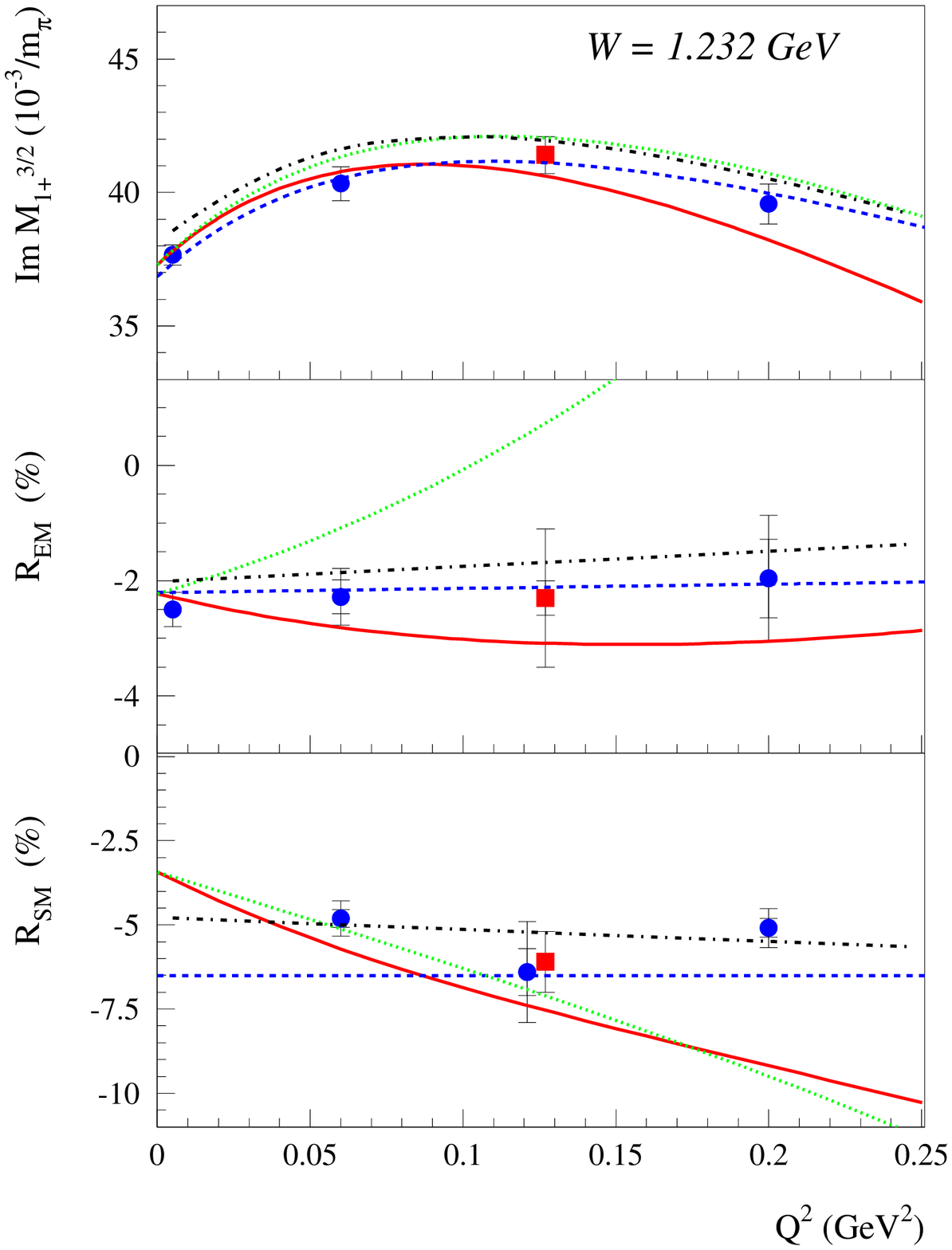}}
\vspace{-7.cm}
\rightline{\includegraphics[height=6.5cm]{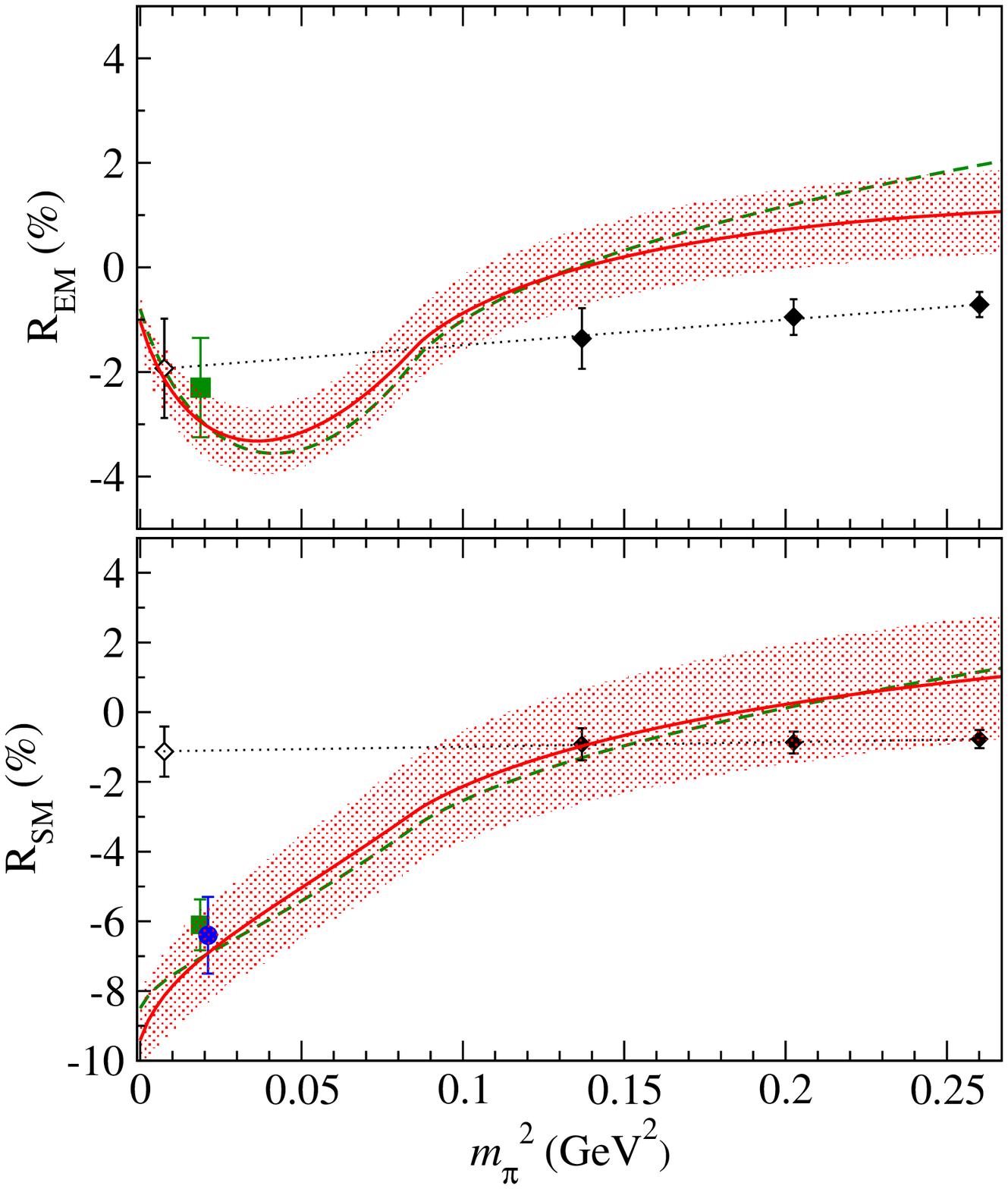}}
\caption{
Left panel : resonant multipoles
of pion electroproduction as function $Q^2$ at the $\Delta$-resonance position.
Dotted curves : $\Delta$-contribution alone.  
Solid curves : results of the NLO calculation of~\cite{Pascalutsa:2005}. 
Also shown are results of the 
SAID analysis (FA04K)~\cite{Arndt:2002xv} (dashed-dotted curves),
and  the MAID 2003 analysis~\cite{MAID98} (dashed curves).
The data points are from BATES (red solid square) and 
MAMI (blue solid circles), for references see~\cite{Pascalutsa:2006up}.
Right panel : pion mass dependence of
$E2/M1$ ($R_{EM}$) (upper panel) and $C2/M1$ ($R_{SM}$) (lower panel), 
at $Q^2=0.1$ GeV$^2$.
Data points from MAMI (blue circle) and BATES (green squares). 
The three filled black diamonds at larger $m_\pi$   
are lattice calculations~\protect\cite{Alexandrou:2004xn}, 
whereas the open diamond near $m_\pi \simeq 0$  
represents their extrapolation assuming linear dependence in $m_\pi^2$. 
The red solid curves are the NLO result of ~\cite{Pascalutsa:2005}, and   
the error bands are an estimate of the theoretical uncertainty. 
}
\label{ratios}
\end{figure}

In Fig.~\ref{ratios} we show the next-to-leading order (NLO) $\chi$EFT results
for the $Q^2$ dependence of the $\gamma N \Delta$ resonant multipoles, 
at the resonance position, characterized by magnetic dipole ($M1$), 
electric quadrupole ($E2$) and Coulomb quadrupole ($C2$) transitions.  
The red solid curve is {\it with} and the green dotted 
curve {\it without} the chiral-loop corrections. 
We observe from the figure that the chiral loops play a crucial role in
the low momentum-transfer dependence of the $E2/M1$ ($R_{EM}$) ratio. 
The effect of the ``pion cloud'' is most 
pronounced in  the $E2$ $\gamma N\Delta$ transition. 

Such $\chi$EFT was found to be useful in a dual
way~: both to describe observables and as a tool to extrapolate lattice
results. The latter is demonstrated on the right panel of Fig.~\ref{ratios} 
which shows the $m_\pi$-dependence of the ratios 
$R_{EM}$ and $R_{SM}$ within the $\chi$EFT 
framework in comparison with the lattice QCD calculations.  
It is seen in particular that 
the opening of the decay channel due to $\Delta \to \pi N$ decay, 
at $m_\pi = M_\Delta - M_N$, leads to 
strong non-analytic behavior in quark mass. For lattice studies, 
the $\Delta(1232)$ resonance, as a purely elastic resonance, 
is an ideal test case to study the issues, due to opening of decay channels, 
which are relevant for the whole baryonic spectrum.

\section{Conclusions}
\label{sec5}

The recent round of double polarization measurements in elastic $eN$
scattering have led to precise extractions of nucleon e.m. FFs. 
At low $Q^2$ values, they allow to test the pion cloud of the nucleon. 
At $Q^2$ values of several GeV$^2$, the difference between the new
polarization data and the unpolarized Rosenbluth data is now mainly understood 
as being due to two-photon exchange effects which largely contribute to 
the Rosenbluth data in contrast to the polarization transfer results. 
The precise new data base for the nucleon e.m. 
FFs allows to map out the transverse quark densities in a fast moving nucleon. 
On the theoretical side, the state-of-the-art in 
ab initio calculations of nucleon e.m. FFs from 
lattice QCD is that unquenched calculations are now feasible for the isovector 
nucleon e.m. FFs. They are performed for pion masses down to 
about $m_\pi \simeq 350$~MeV, into the regime where chiral effects are
important. Although the lattice isovector FFs show clear tendency to 
approach their experimental values as the pion mass decreases to its physical 
value, the precision of the FF data clearly calls for lattice 
calculations closer to the chiral limit. Besides, for a calculation of
isoscalar FFs, allowing the separate predictions of proton and neutron
e.m. FFs, the evaluation of disconnected diagrams is a high priority. 

GPDs have emerged over the past decade as a unifying theme in hadron physics 
linking the spatial densities extracted from FFs with the quark momentum 
distribution information residing in quark structure functions. 
They also allow an access to the angular momentum carried by quarks 
( and gluons ) in the nucleon. 
The first round of experiments dedicated to measure 
hard exclusive processes indicate the dominance 
of the twist-2 mechanism at available momentum transfers. 
Future programs at Compass and a uniquely dedicated program at JLab@12GeV 
will provide a wide survey of such distributions. 

The recent results on the e.m. excitation 
of the $\Delta(1232)$ resonance, 
have also briefly been reviewed.  
Recently developed chiral effective field theories 
have emerged which can be useful both to describe observables 
and as a tool to extrapolate lattice QCD results.

\end{document}